\documentstyle[12pt]{article}
\def\beq{\begin{equation}}
\def\eeq{\end{equation}}
\def\bea{\begin{eqnarray}}
\def\eea{\end{eqnarray}}

\def\ba{\begin{array}}
\def\ea{\end{array}}

\def\,{\"{U}}
\def\6{\.{I}}

\begin{document}

\title{Novel Bound States Treatment of the Two Dimensional
Schr\"{o}dinger Equation with Pseudocentral Potentials Plus
Multiparameter Noncentral Potential}

\baselineskip 0.9cm
\author{Metin Akta\c{s}\thanks{Corresponding author: E-mail: metin@karatekin.edu.tr}\\[.5cm]
Department of Physics, Faculty of Science\\
Karatekin University of \c{C}ank{\i}r{\i}, 18100, \c{C}ank{\i}r{\i},
Turkey}

\date{\today}
\maketitle
\normalsize

\begin{abstract}

\noindent By converting the rectangular basis potential $V(x,y)$
into the form as $V(\textrm{r})+V(\textrm{r},\varphi)$ described by
the pseudo central plus noncentral potential, particular solutions
of the two dimensional Schr\"{o}dinger equation in plane-polar
coordinates have been carried out through the analytic approaching
technique of the Nikiforov and Uvarov (NUT). Both the exact bound
state energy spectra and the corresponding bound state wavefunctions
of the complete system are determined explicitly and in closed
forms. Our presented results are identical to those of the previous
works and they may also be useful for investigation and analysis of
structural characteristics in a variety of quantum systems.\\[0.1cm]

\noindent equation, bound states. \\[0.1cm]

\end{abstract}

\newpage

\section{Introduction}

\noindent The pioneering proposal of Hartmann [1] namely as the
ring-shaped potential and its applications [2] has offered much
attention in recent times. The so-called Hartmann ring-shaped
potential $V(\textrm{r},\varphi)$ exhibits a special case for the
noncentral class potentials. It can be expressed as the combinations
of an attractive Coulombic term $(\textrm{A}/\textrm{r})$ and a
repulsive noncentral term
$(\textrm{B}/\textrm{r}^{2}sin^{2}\varphi)$ together with the
parameters $\textrm{A}=2a_{0}\varepsilon_{0}\eta\sigma^{2}$ and
$\textrm{B}=-qa_{0}\varepsilon_{0}\eta^{2}\sigma^{2}$. Here also
$a_{0}=\hbar^{2}/me^{2}$, $\varepsilon_{0}=-me^{4}/2\hbar^{2}$ stand
for the Bohr radius, the ground states energy of the hydrogen atom
respectively with the dimensionless parameters $q$, $\eta$ and
$\sigma$ varying from $1$ up to $10$.\\

\noindent By proving the evidence of `accidental' degeneracies in
quantum levels, the dynamical invariance of algebra for the Hartmann
potential was firstly handled by the authors Kibler, Negadi [3],
Gerry [4] and Kibler, Winternitz [5] respectively. On behalf of the
work [5], they have realized that the dynamical symmetry of quantum
systems gives rise to the `accidental' degeneracy which is exhibited
the discrete spectrum. Zhedanov [6] also proposed the dynamical
group symmetry approach to prove the `hidden' symmetry algebra of
the Hartmann and oscillator types ring-shaped potential systems.\\

\noindent Two distinct forms of noncentral class potentials are
realized in the literature. The first form refers to the Hartmann
type ring-shaped potential. The second form firstly introduced by
Quesne [7] to investigate the `accidental' degeneracy corresponds to
the oscillator type ring-shaped potential. It is employed by
combining the terms of harmonic oscillator $({\sim \textrm{r}^{2}})$
and $(\textrm{r}, \varphi)$-dependent noncentral potential [8]. It
is pointed out that in the first classes the leading term implies
the Coulombic discrete energies though in the second ones the
leading term suggests the harmonic oscillator basis discrete energies.\\

\noindent Quite a few applications have been encountered in such
works on the subject of noncentral potentials (NCPs) from past to
the present time. For example, general relations between the elastic
constants and the central forces in hexagonal materials as well as
noncentral forces in $\textit{fcc}$ monatomic structures were given
by Johnson [9]. Some certain structures with these type potentials
were also studied by the authors [10-16]. Ermakov type invariants
were also discussed with respect to them by Makowski in exact
classical limit of quantum mechanics [17]. Besides, the derivations
were achieved for determining the stress and elastic constants in
systems of particles via noncentral two-body potentials by Murat and
Kantor [18]. Several authors has employed them for the aspect of
scattering analysis in continuum bound states  [19-25]. Fredholm
theory was applied to the Lippmann-Schwinger equation and the
generalized Levinson theorem was also proved for NCPs [19]. Forward
scattering in a system with those potentials was discussed
by [20].\\

\noindent Recently, much considerable effort for variety forms of
NCPs has been expanded on the solutions of Schr\"{o}dinger, Dirac
and Klein-Gordon equations. The Feynman's path integral treatment
[26-30] and the Green's function technique [31, 32], the (Lie)
algebraic/group theoretical approach [33-36], nonbijective canonical
transformation [3, 37], supersymmetric (SUSY) quantum mechanical
formalism [38-46] and the NU-analytic method [47-58] as well as the
applications for both relativistic [59-71] and other nonrelativistic
[72-82] cases are available in the literature.\\

\noindent Due to the above reasons, NCPs may be helpful for
determining the structural properties of systems in such cases
(\textit{e.g.} elasticity and stress factors, point-defects of
surfaces etc.) in variety fields of physics and chemistry. The
ring-shaped like structures, \textit{i.e.}, cyclic polyenes, benzene
and benzene-like structures (\textit{e.g.} graphene), and
interactions between deformed pair of nuclei are good examples for the discussion.\\

\noindent The key idea `noncentrality concept' of all potentials
plays a privilege role to provide the extensive solutions for
concerning quantum systems. That is, if one of NCPs preserves the
separability condition for achieving the solutions of system, it can
therefore be split into two parts for such cases of radial and
angular dependent equations that they may readily be solved.\\

\noindent Our primary goal of this work is to present the
expressions of the exact bound states both the energy spectra and
the wavefunctions of the two dimensional Schr\"{o}dinger equation by
proposing two definite potential forms in which they are called the
pseudocentral potentials plus multiparameter NCP. It is the author's
hope to further contribute to applying our approach and results for
the most recent quantum systems and for exactly solvable systems
with centrally style potentials [83, 84].  It may also be applicable
for conforming the initial boundary-value problems of
certain physical systems (\textit{e.g.} Dirichlet, Neumann etc.) in two-dimension [85-87].\\

\noindent The organization scheme of the study is given as follows:
Section 2 is devoted to construction of the two different
rectangular potentials and converting of them to be solvable form
for the Schr\"{o}dinger equation in plane polar coordinates. Section
3 covers the solutions of separated equations via the NU-method.
Section 4 is responsible for dealing with the conclusions and
remarks.

\section{Potential Cases and Separation of Variables}

\noindent Here we will firstly describe the essential steps of
converting the rectangular basis potentials as $V_{I}(x,y)$ and
$V_{II}(x,y)$ into the plane polar forms as
$V_{I}(\textrm{r})+V_{I}(\textrm{r},\varphi)$ and
$V_{II}(\textrm{r})+V_{II}(\textrm{r},\varphi)$. Secondly, by
applying the separational procedure to the two dimensional
Schr\"{o}dinger equation (2-dim SE) then we are going to
deal with the solutions of separated equations.\\

\noindent We propose the rectangular form physical potentials as
follows

\begin{equation}\label{1}
    V_{I}(x,y)=\left(\frac{A_{0}}{\sqrt{x^{2}+y^{2}}}+B_{0}\right)^{2}
    +\frac{1}{x^{2}y^{2}}\left[\frac{Bx^{4}+Cy^{4}}{(x^{2}+y^{2})}+Dx^{2}+Fy^{2}+G(x^{2}+y^{2})\right]
\end{equation}\\

\noindent and

\begin{equation}\label{2}
    V_{II}(x,y)=\left(A_{1}\sqrt{x^{2}+y^{2}}+\frac{B_{1}}{\sqrt{x^{2}+y^{2}}}\right)^{2}
    +\frac{1}{x^{2}y^{2}}\left[\frac{Bx^{4}+Cy^{4}}{(x^{2}+y^{2})}+Dx^{2}+Fy^{2}+G(x^{2}+y^{2})\right]
\end{equation}\\

\noindent The squared leading terms in equations (1) and (2) refer
to the pseudo Coulombic/modified Kratzer and the pseudoharmonic
circular oscillator potentials as well. The rest parts of them
correspond to the equivalent multiparameter potentials which will be
converted into the noncentral case.\\

\noindent Let us consider the cartesian coordinate transformations
as [88] $x=\textrm{r}\cos\varphi$, $y=\textrm{r}\sin\varphi$,
$\varphi=\tan^{-1}(y/x)$ and $\textrm{r}=\sqrt{x^{2}+y^{2}}$, by
putting these into the above equations then they are readily
converted into the plane polar forms as

\begin{eqnarray}\label{3}
    V(\textrm{\textbf{r}})=\nonumber V_{I}(\textrm{r})+V_{I}(\textrm{r}, \varphi)
    =\left(\kappa_{0}+\frac{\kappa_{1}}{\textrm{r}}
    +\frac{\kappa_{2}}{\textrm{r}^{2}}\right)\nonumber
    &+&\frac{1}{\textrm{r}^{2}}
    \left\{B\cot^{2}\varphi+C\tan^{2}\varphi+D\csc^{2}\varphi\right. \nonumber\\[0.3cm]
    &+&\left.
    F\sec^{2}\varphi+G\sec^{2}\varphi\csc^{2}\varphi\right\}
\end{eqnarray}\\

\noindent with the constants $\kappa_{0}=B_{0}^{2}=D_{e}$,
$\kappa_{1}=2A_{0}B_{0}=2D_{e}r_{e}$ and
$\kappa_{2}=A_{0}^{2}=D_{e}r_{e}^{2}$ as in Ref. [50]. The
consecutive terms, in parenthesis, refer to the pseudocentral
potential ($PCP1$) called as the pseudo Coulombic/modified Kratzer
case and the rest part constitutes the multiparameter noncentral
potential labelled as $(mp-NCP)$. By applying the conversion process just only the equation (2) gives\\

\begin{equation}\label{4}
    V(\textrm{\textbf{r}})=V_{II}(\textrm{r})+V_{II}(\textrm{r}, \varphi)=
    V_{0}+A_{1}^{2}~\textrm{r}^{2}+\frac{B_{1}^{2}}{\textrm{r}^{2}}
    +(mp-NCP)
\end{equation}\\

\noindent where we have used the notations
$V_{0}=2A_{1}B_{1}=(\kappa/4)r_{0}^{2}$~ with
$A_{1}=(\kappa/8)^{1/2}$,~ and $B_{1}=(\kappa/8)^{1/2}~r_{0}^{2}$
~as [84]. The consecutive three-term, in equation (4), establishes
the pseudocentral harmonic oscillator potential ($PCP2$).\\

\noindent Let us now regard the two dimensional rectangular basis
Schr\"{o}dinger equation (SE). It can be written

\begin{equation}\label{5}
    \frac{\partial^{2}\Psi}{\partial x^{2}}+\frac{\partial^{2}\Psi}{\partial y^{2}}
    +\frac{2m}{\hbar^{2}}\left[E-V(x,y)\right]\Psi(x,y)=0.
\end{equation}\\

\noindent By employing the task of the polar coordinate
transformation on equation (5), we rewrite it as\\

\begin{equation}\label{6}
    \left[\frac{1}{\textrm{r}}\frac{\partial}{\partial \textrm{r}}\left(\textrm{r}\frac{\partial}
    {\partial \textrm{r}}\right)+\frac{1}{\textrm{r}^{2}}\frac{\partial^{2}}{\partial\varphi^{2}}+\frac{2m}
    {\hbar^{2}}[E-V(\textrm{\textbf{r}})]\right]\Psi(\textrm{r}, \varphi)=0.
\end{equation}\\

\noindent The separation of variables procedure cannot be carried
out when rectangular coordinates are employed since the potential
energies (1)and (2) are functions of $(x^{2}+y^{2})^{-1}$ and
$(x^{2}+y^{2})^{-1/2}$. Due to the fact that they are not split into
terms, it is required changing to the plane polar coordinates.\\

\noindent Using the ansatz wavefunction in equation (6) for
successively concerning the equations (3) and (4),

\begin{equation}\label{7}
    \Psi(\textrm{r},
    \varphi)=\frac{1}{\sqrt{\textrm{r}}}\textsl{U}(\textrm{r})\Phi(\varphi)
\end{equation}\\

\noindent leads to the following equations in which two of them are
radial-dependent correspond to the first and second classes of NCPs

\begin{equation}\label{8}
    \frac{d^{2}\textsl{U}_{A}}{d\textrm{r}^{2}}+\left(\widetilde{\textsc{E}}_{HRS}-\frac{\Lambda_{0}}{\textrm{r}}
    -\frac{\Lambda}{\textrm{r}^{2}}\right)\textsl{U}_{A}(\textrm{r})=0,
\end{equation}\\

\noindent and\\

\begin{equation}\label{8}
    \frac{d^{2}\textsl{U}_{B}}{d\textrm{r}^{2}}+\left(\widetilde{\textsc{E}}_{RSO}-\widetilde{A}\textrm{r}^{2}
    -\frac{\Gamma}{\textrm{r}^{2}}\right)\textsl{U}_{B}(\textrm{r})=0,
\end{equation}\\

\noindent and the angle-dependent equation

\begin{equation}\label{10}
    \frac{d^{2}\Phi}{d\varphi^{2}}+\left\{\emph{\textbf{M}}^{2}-\left[
    \left(\frac{\bar{D}+\bar{B}\cos^{2}\varphi}{\sin^{2}\varphi}\right)
    +\left(\frac{\bar{F}+\bar{C}\sin^{2}\varphi}{\cos^{2}\varphi}\right)
    +\left(\frac{\bar{G}}{\sin^{2}\varphi\cos^{2}\varphi}\right)\right]\right\}\Phi(\varphi)=0.
\end{equation}

\noindent In these equations,  we use the specifications
$\widetilde{\textsc{E}}_{HRS}=2m(\textsc{E}-\kappa_{0})/\hbar^{2}$,
$\Lambda_{0}=(2m\kappa_{1})/\hbar^{2}$ and
$\Lambda=[\bar{\kappa}_{2}+(\emph{\textbf{M}}^{2}-\frac{1}{4})]$
with
$\bar{\kappa}_{2}=(2m\kappa_{2})/\hbar^{2}=(2mA_{0}^{2})/\hbar^{2}$;
$\widetilde{\textsc{E}}_{RSO}=2m(\textsc{E}-V_{0})/\hbar^{2}$,
$\widetilde{A}=(2mA_{1}^{2})/\hbar^{2}$, and
$\Gamma=[\bar{B}_{1}+(\emph{\textbf{M}}^{2}-\frac{1}{4})]$ with
$\bar{B}_{1}=(2mB_{1}^{2}/\hbar^{2})$. In addition,
$\widetilde{\textsc{E}}_{HRS}$ and  $\widetilde{\textsc{E}}_{RSO}$
belong to the energies of the Hartmann ring-shaped $(HRS)$ and the ring-shaped oscillator $(RSO)$ potentials, respectively.\\

\noindent Besides we label the energy parameters for noncentral
potential part as $\bar{B}=(2mB)/\hbar^{2}$,
$\bar{C}=(2mC)/\hbar^{2}$, $\bar{D}=(2mD)/\hbar^{2}$,
$\bar{F}=(2mF)/\hbar^{2}$ and $\bar{G}=(2mG)/\hbar^{2}$ with the
separation constant $\emph{\textbf{M}}$ recognized as the angular
momentum.

\section{Achieving the Solutions of Separated Equations}

\subsection{Solution of the $\varphi-$Angle Dependent Equation}

\noindent Let us begin firstly to examine some solutions of the
equation (10). By introducing a transformation as
$t=\sin^{2}\varphi$ hence the transformed hypergeometric equation
holds a form as

\begin{equation}\label{11}
    \frac{d^{2}\Phi}{dt^{2}}+\frac{(1-2t)}{2t(1-t)}\frac{d\Phi}{dt}
    +\frac{1}{[2t(1-t)]^{2}}\left(\alpha t^{2}+\beta t+\gamma\right)\Phi(t)=0,
\end{equation}\\

\noindent where the parameters are
$\alpha=(-\emph{\textbf{M}}^{2}+\bar{B}-\bar{C})$,
$\beta=(\emph{\textbf{M}}^{2}-\bar{D}-\bar{F})$ and
$\gamma=-(\bar{B}+\bar{D}+\bar{G})$. By following [47], a
hypergeometric equation is described by\\

\begin{equation}\label{12}
    \frac{d^{2}u}{dt^{2}}+\frac{\tilde{\tau}(t)}{\sigma(t)}\frac{d u}{dt}
    +\frac{\tilde{\sigma}(t)}{{\sigma}^{2}(t)}u(t)=0,
\end{equation}\\

\noindent then comparing the equations (11) and (12)
term-by-term allow us to write the polynomials\\

\begin{equation}\label{13}
\tilde{\tau}(t)=(1-2t)\qquad \sigma(t)=2t(1-t)\qquad and \qquad
\tilde{\sigma}(t)=(\alpha t^{2}+\beta t+\gamma).
\end{equation}\\

\noindent They are all used in solving of the quadratic equation\\

\begin{equation}\label{14}
     \pi(t)=\frac{1}{2}\left(\sigma^{\prime}-\tilde{\tau}\right)\pm\frac{1}{2}\sqrt{\left(\sigma^{\prime}
     -\tilde{\tau}\right)^{2}+4(k\sigma-\tilde{\sigma})},
\end{equation}\\

\noindent where the prime factor of $\sigma$ denotes the
differential at first degree. In this equation the determination of
$k$ is very essential step for the calculation of $\pi(t)$ in which
the inner part of the square root is required to become the square
form with respect to the polynomials.\\

\noindent The substitution process of polynomials for $\pi(t)$ and
the arrangement of it for $k$ provide us\\

\begin{equation}\label{15}
     \pi(t)=\frac{1}{2}(1-2t)\pm\frac{1}{2}\sqrt{a_{1}t^{2}+a_{2}t+a_{3}}
\end{equation}\\

\noindent with the parameters $a_{1}=(-4\alpha-8k+4)$,
$a_{2}=(-4\beta+8k-4)$ and $a_{3}=(1-4\gamma)$. The values of $k$
which bring about the releasing from the square root of $\pi(t)$ can
be determined by solving the quadratic equation

\begin{equation}\label{16}
     k^{2}+\left(-\beta-2\gamma-\frac{1}{2}\right)k+\frac{1}{4}\left({\beta^{2}}+2\beta
     +\alpha+4\gamma(1-\alpha)\right)=0.
\end{equation}\\

\noindent Then it yields the double roots as follows\\

\begin{equation}\label{17}
     k_{1,2}=\left(\gamma+\frac{\beta}{2}+\frac{1}{4}\right)\pm\frac{1}{4}
     \sqrt{\left(1-4\gamma\right)\left\{1-4\left[\gamma+(\alpha+\beta)\right]\right\}}.
\end{equation}\\

\noindent By inserting these roots into the equation (15) leads to
the fourfold roots for $\pi(t)$
as\\

\begin{equation}\label{18}
     \pi_{1,2}(t)=\frac{1}{2}(1-2t)\pm\frac{1}{4}\left\{\left[\sqrt{1-4\gamma}
     -\sqrt{1-4\left[\gamma+(\alpha+\beta)\right]}\right]t-\sqrt{1-4\gamma}\right\}
\end{equation}\\

\noindent for negative root of $k=k_{1}$ and\\

\begin{equation}\label{19}
     \pi_{3,4}(t)=\frac{1}{2}(1-2t)\pm\frac{1}{4}\left\{\left[\sqrt{1-4\gamma}
     +\sqrt{1-4\left[\gamma+(\alpha+\beta)\right]}\right]t-\sqrt{1-4\gamma}\right\}
\end{equation}\\

\noindent for positive root of $k=k_{2}$, respectively. The
resulting equations (13), (17), (18) and (19) play a key role to
establish both the energy spectra and the wavefunctions of the
system.

\subsection{Bound States Eigenvalues}

\noindent Now we can put forward the procedure for determining the
energy spectra of the systems. For the purpose of this, we again
follow [47] and a novel form of the energy equation is adapted as

\begin{eqnarray}\label{20}
    \lambda\equiv\lambda_{n}&=&k+\frac{d\pi}{dt}\nonumber\\[0.2cm]
    &=&-n\left(\frac{d}{dt}(\tilde{\tau}+2\pi)-2n+2\right).
\end{eqnarray}\\

\noindent Each root $k_{1}$ for $\pi_{1,2}(t)$ and $k_{2}$ for
$\pi_{3,4}(t)$ can be applied which one of them is satisfied the
condition $\tau^{\prime}<0$
$[\tau^{\prime}=\frac{d}{dt}(\tilde{\tau}+2\pi)]$. Possible
solutions have often been carried out by the negative roots of the
functions $\pi(t)$. Hence we accomplish the result of the angular
momentum $\emph{\textbf{M}}$ which implies the reasonable meaning
for physically

\begin{eqnarray}\label{21}
    \emph{\textbf{M}}^{2}&=&[\bar{D}+\bar{F}+2(\bar{B}+\bar{G})]+\frac{1}{2}\left(-1
    +\sqrt{[1+4(\bar{B}+\bar{D}+\bar{G})][1+4(\bar{C}+\bar{F}+\bar{G})]}\right)\nonumber\\
    &+&\left\{(2n_{0}+1)\left[(2n_{0}+1)+\frac{1}{2}(\sqrt{1+4(\bar{B}+\bar{D}+\bar{G})}
    +\sqrt{1+4(\bar{C}+\bar{F}+\bar{G})})\right]+1\right\}
\end{eqnarray}

\noindent with $n_{0}=0,1,2\ldots$\\

\noindent One of our basic interests in this step is to show and
present how to get the energy eigenvalue results for the
radial-dependent equations (8) and (9) with the potential functions
(3) and (4) respectively by employing the analogy procedure. It is
obvious that the equation (38) and its converted equation (50) given
as [48] are identical form to the equation (8). Accordingly, it
should also admit to be similar form solutions. Following the same
procedure as in [48] and accepting the energy parameters of [50]
give rise to

\begin{eqnarray}\label{22}
    \textsc{E}_{n_{1}}&=&\kappa_{0}-\frac{\hbar^{2}\Lambda_{0}^{2}}{8m}\left[\left(n_{1}+\frac{1}{2}\right)
    +\frac{1}{2}\sqrt{1+4\Lambda}\right]^{-2}\nonumber\\[0.2cm]
    &=&D_{e}-\frac{2m}{\hbar^{2}}D_{e}^{2}r_{e}^{2}\left[\left(n_{1}+\frac{1}{2}\right)+\frac{1}{2}
    \sqrt{\frac{2mD_{e}r_{e}^{2}}{\hbar^{2}}+\emph{\textbf{M}}^{2}}~\right]^{-2}, \qquad
    n_{1}=0,1,\ldots
\end{eqnarray}\\

\noindent Note that the equation (57) in [48] and the energy
expression (22) referred to (PCP1) plus $(mp-NCP)$ potential exhibit
similar characteristic by virtue of the presence of the
Coulombic term in potential energy.\\

\noindent The energy spectra expression for $(PCP2)$ plus $(mp-NCP)$
can also be determined analogously by dealing with [56] and [84].
The equation (15) and its converted form (18) in [56] hold similar
structure to the form of the equation given by (9). Following the
procedure as [56] step-by-step and the arrangement of the result
allow us to write analogously

\begin{eqnarray}\label{23}
    \textsc{E}_{n_{2}}&=&V_{0}-\hbar\omega\left[(2n_{2}+1)+\frac{1}{2}\sqrt{1+4\Gamma}\right]\nonumber\\[0.2cm]
    &=&\frac{1}{8}\kappa r_{0}^{2}-\frac{\hbar}{2}\sqrt{\frac{\kappa}{m}}\left[(2n_{2}+1)
    +\sqrt{\left(\frac{m}{4\hbar^{2}}\right)\kappa r_{0}^{4}+\emph{\textbf{M}}^{2}}~\right], \qquad
    n_{2}=0,1,\ldots
\end{eqnarray}\\

\noindent where $\kappa=4m\omega^{2}$ with
$\omega=\sqrt{(2mA_{1}^{2})/\hbar^{2}}$ and the angular momentum
$\emph{\textbf{M}}$.

\subsection{Bound State Wavefunctions}

\noindent The essential steps for performing a set of wavefunction
for the complete system should now be examined. In this point we are
going to introduce the proper notations in the solutions of
$k=k_{2}=\gamma+\frac{\beta}{2}+\frac{1}{4}\left(\sqrt{(1+\beta_{1})(1+\beta_{2})}
~\right)$ for which this is the correspondence polynomial of
$\pi(t)=\pi_{4}(t)=\frac{1}{2}(1-2t)-\frac{1}{4}\left\{\left[\sqrt{1+\beta_{1}}
+\sqrt{1+\beta_{2}}\right]t-\sqrt{1+\beta_{1}}\right\}$ with
$\beta_{1}=-4\gamma$ and $\beta_{2}=-4[\gamma+(\alpha+\beta)]$ as
well as $\alpha$, $\beta$ and $\gamma$ are given in the equation (11).\\

\noindent Let us start to write the equation

\begin{equation}\label{24}
    \frac{\pi(t)}{\sigma(t)}=\frac{d}{dt}[\ln\phi(t)]
\end{equation}\\

\noindent and from the equation (13), straightforward calculations
give for

\begin{equation}\label{25}
    \phi(t)=[t(1-t)]^{\delta/2}
\end{equation}\\

\noindent where
$\delta=[1+(\sqrt{1+\beta_{1}}+\sqrt{1+\beta_{2}}~)/4]$. The weight
function [47] $\varrho(t)$ is determined by

\begin{equation}\label{26}
    \frac{d}{dt}(\varrho\sigma)=\varrho\tau
\end{equation}\\

\noindent with the expression $\tau=\tilde{\tau}+2\pi$. The
calculating procedure yields

\begin{equation}\label{27}
    \varrho(t)=[t(1-t)]^{\delta-1}
\end{equation}\\

\noindent where
$\delta-1=(\sqrt{1+\beta_{1}}+\sqrt{1+\beta_{2}}~)/4$. Then the
Rodriguez formula is

\begin{equation}\label{28}
    y_{n}(t)=\frac{C_{n}}{\varrho(t)}\frac{d^{n}}{dt^{n}}[\sigma^{n}(t)\varrho(t)],
\end{equation}\\

\noindent where $C_{n}$ is the normalization constant. Use of
$\varrho(t)$ and $\sigma(t)$ provides\\

\begin{equation}\label{29}
    y_{n}(t)=\bar{C}_{n}~\textsc{P}_{n}^{(\mu_{1},~\mu_{2})}(t),\qquad
    t=\sin^{2}\varphi
\end{equation}\\

\noindent with $\bar{C}_{n}=2^{n}C_{n}$,
$\mu_{1}=\sqrt{1+\beta_{1}}$ and $\mu_{2}=\sqrt{1+\beta_{2}}$ as
well as $\textsc{P}_{n}^{(\mu_{1},~\mu_{2})}(t)$ stands for the
Jacobi polynomial. The net wave function can thus be written as\\

\begin{eqnarray}\label{30}
    \Phi(t)&=&\phi(t)~y_{n}(t)\nonumber\\[0.2cm]
    &=&\bar{C}_{{n}_{0}}~[t(1-t)]^{\delta/2}~\textsc{P}_{{n}_{0}}^{(\mu_{1},~\mu_{2})}(t)
\end{eqnarray}\\

\noindent In order to construct the wavefunctions for the
radial-dependent equations (8) and (9), we should consider the
equations from (58) to (62) as [48] and the equations from (35) to
(42) as [56], respectively. By means of the analogy procedure one
gets

\begin{eqnarray}\label{31}
    \textsl{U}(\textrm{r})&\equiv&\textsl{U}_{A}(\textrm{r})=\phi(\textrm{r})~y_{n}(\textrm{r})\nonumber\\[0.2cm]
    &=&\bar{C}_{n_{1}}~\textrm{r}^{p}~e^{-\sqrt{\textsc{E}_{HRS}}~\textrm{r}}~\textsc{L}_{{n}_{1}}^{p}(\textrm{r})
\end{eqnarray}

\noindent and

\begin{eqnarray}\label{32}
    \textsl{U}(\textrm{s})&\equiv&\textsl{U}_{B}(\textrm{s})=\phi(\textrm{s})~y_{n}(\textrm{s})\nonumber\\[0.2cm]
    &=&\bar{C}_{n_{2}}~s^{\zeta/4}~e^{-\frac{1}{2}\sqrt{\tilde{A}}~\textrm{s}}~\textsc{L}_{n_{2}}^{q}(\textrm{s}),\qquad
    s=\textrm{r}^{2}
\end{eqnarray}\\
\noindent where we have used the short notations
$p=1+2\sqrt{1+4\Lambda}$, $\zeta=1+\sqrt{1+4\Gamma}$ and
$q=(\zeta-1)/2$. Also $\textsc{L}_{n_{1}}^{p}(\textrm{r})$ and
$\textsc{L}_{n_{2}}^{q}(\textrm{s})$ stand for the Laguerre
polynomials, respectively.

\subsection{Complete Solutions of Bound Wavefunctions}

\noindent In this part it is required to gather all wavefunctions in
which they are satisfied by the equations of (8), (9) and (10) with
respect to the ansatz wavefunction (7). If we substitute the
equations (30) and (31) into the equation (7) one yields\\
\begin{equation}\label{33}
    \Psi(\textrm{r}, \varphi)=
    \bar{N}~\textrm{r}^{(p-\frac{1}{2})}~e^{-\sqrt{\textsc{E}_{HRS}}~\textrm{r}}~\textsc{L}_{{n}_{1}}^{p}(\textrm{r})~
    \left(\sin2\varphi\right)^{2\delta}
    ~\textsc{P}_{{n}_{0}}^{(\mu_{1},~\mu_{2})}(\sin^{2}\varphi)
\end{equation}\\
\noindent where
$\bar{N}=(\bar{C}_{n_{0}}\cdot\bar{C}_{n_{1}})/2^{\delta}$,
$\sin2\varphi=2\sin\varphi\cos\varphi$ and $n_{0},~n_{1}=0, 1, 2
\ldots$. By employing the cartesian coordinate transformations then
it is transformed
into\\
\begin{eqnarray}\label{34}
    \Psi(x, y)&=&\bar{N}~(x^{2}+y^{2})^{(p-\frac{1}{2})/2}~e^{-\sqrt{\textsc{E}_{HRS}}~\sqrt{x^{2}+y^{2}}}
    ~[x y/(x^{2}+y^{2})]^{2\delta}\nonumber\\[0.3cm]
    &\times&\textsc{L}_{{n}_{1}}^{p}[(x^{2}+y^{2})^{1/2}]~
    \textsc{P}_{{n}_{0}}^{(\mu_{1},~\mu_{2})}[y^{2}/(x^{2}+y^{2})].
\end{eqnarray}\\
\noindent In this time we consider the equations (30) and (32) for
ansatz equation (7) one gets\\
\begin{equation}\label{35}
    \Psi(\textrm{r}, \varphi)=
    \tilde{N}~\textrm{r}^{(\zeta-1)/2}~e^{-\frac{1}{2}\sqrt{\tilde{A}}~\textrm{r}^{2}}~\textsc{L}_{{n}_{2}}^{q}(\textrm{r}^{2})~
    \left(\sin2\varphi\right)^{2\delta}
    ~\textsc{P}_{{n}_{0}}^{(\mu_{1},~\mu_{2})}(\sin^{2}\varphi)
\end{equation}\\
\noindent with $n_{0},~n_{2}=0, 1, 2 \ldots $. Consequently, it can
be converted into\\
\begin{eqnarray}\label{36}
    \Psi(x, y)&=&\tilde{N}~(x^{2}+y^{2})^{(\zeta-1)/4}~e^{-\frac{1}{2}\sqrt{\tilde{A}}~(x^{2}+y^{2})}
    ~[x y/(x^{2}+y^{2})]^{2\delta}\nonumber\\[0.3cm]
    &\times&\textsc{L}_{{n}_{2}}^{q}(x^{2}+y^{2})~
    \textsc{P}_{{n}_{0}}^{(\mu_{1},~\mu_{2})}[y^{2}/(x^{2}+y^{2})],
\end{eqnarray}\\
\noindent where
$\tilde{N}=(\bar{C}_{n_{0}}\cdot\bar{C}_{n_{2}})/2^{\delta}$.

\section{Conclusions and Remarks}

\noindent In this study we have performed the exact analytical bound
state solutions both the energy spectra and the corresponding
wavefunctions for the two dimensional Schr\"{o}dinger equation in
plane polar coordinates. The rectangular basis potentials proposed
by the equations (1) and (2) are converted into the planar forms in
solving of the 2-dim SE through the NU-analytic technique.\\

\noindent Let us remark that the discussion of noncentral class
potentials for such quantum systems is limited in the literature.
This is the main reason of all not satisfying the condition of
analytical solvability. On the other hand, the radial-angular
dependency of such potentials will provide us for examining the
structural properties of planar quantum systems, $\textit{i. e.}$
elasticity, stress factor and point-defects of surfaces etc. It is
expected that our straightforward approach may therefore offers a
solution or provide a model for other identical systems described by [84].\\

\noindent With regard to a variety of potential parameters, possible
cases are briefly examined for energy expressions presented by (22)
and (23). Certain examples are given as follows:\\

\noindent Letting $B=C=D=F=G=0$ the angular-dependent equation (10)
and the equation (21) are ignored, therefore this just refers to the
energies of the pseudo Coulombic potential (PCP1) and the
pseudoharmonic oscillator potential (PCP2). For $D\neq 0$, and
$B=C=F=G=0$, we accomplish the energy eigenvalue results of the
pseudo Coulombic type Hartmann ring-shaped potential and the
pseudoharmonic oscillator type ring-shaped potential as well.\\

\noindent As a final remark we conclude that the two dimensional
angular momentum operator defined as [88]\\

\begin{displaymath}
\hspace{0.01cm} L\Psi=-i\hbar \frac{\partial\Psi}{\partial\varphi}
\end{displaymath}\\
\noindent with the eigenvalue $(\hbar\emph{\textbf{M}})$\\
\begin{displaymath}
\hspace{0.01cm} L\Psi=(\hbar\emph{\textbf{M}})\Psi
\end{displaymath}\\
\noindent where the $\Psi$ corresponds to the complete solutions of
bound wavefunctions (33) and (35), angular momentum expression of
our proposal system is also defined by (21).

\newpage

\noindent $\bf{Acknowledgment}$\\

\noindent The Turkish Scientific and Technological Research Council
(T\"{U}B\.{I}TAK) partially supports this work.

\end{document}